\documentclass[prb,twocolumn,showpacs,floatfix,superscriptaddress]{revtex4}
\usepackage{epsfig,amssymb,amsmath,subfigure}

\newcommand*{\atow}[0]{$\alpha$$\to$$\omega$}
\newcommand*{\eps}[0]{\varepsilon}
\newcommand*{\vasp}[0]{\textsc{vasp}}

\newcommand*{\wien}[0]{\textsc{wien97}}
\newcommand*{\EF}[0]{E_{\text{F}}}
\newcommand*{\Rmin}[0]{R_{\text{min}}}
\newcommand*{\Rmt}[0]{R_{\text{MT}}}
\newcommand*{\Kmax}[0]{K_{\text{max}}}

\newcommand*{\et}[0]{\textit{et~al.}}

\newcommand*{\x}{\times}
\newcommand*{\EE}[1]{\times 10^{{#1}}}

\newcommand*{\be}[0]{\begin{equation}}
\newcommand*{\ee}[0]{\end{equation}}
\newcommand*{\beu}[0]{\begin{equation*}}
\newcommand*{\eeu}[0]{\end{equation*}}
\newcommand*{\ba}[0]{\begin{array}}
\newcommand*{\ea}[0]{\end{array}}

\newcommand*{\entry}[1]{\multicolumn{1}{c}{\underbar{#1}}}
\newcommand*{\rentry}[1]{\entry{\raisebox{1pt}[\depth][\width]{#1}}}

\begin{document}

\title{An Empirical Tight-Binding Model for Titanium Phase Transformations}

\author{D. R. Trinkle}
\affiliation{Materials and Manufacturing Directorate, Air Force Research Laboratory, Wright Patterson Air Force Base, Dayton, Ohio 45433-7817}
\affiliation{Theoretical Division, Los Alamos National Laboratory, Los Alamos, NM 87545}
\author{M. D. Jones}
\affiliation{State University of New York, Buffalo, NY 14260}
\affiliation{Theoretical Division, Los Alamos National Laboratory, Los Alamos, NM 87545}
\author{R. G. Hennig}
\affiliation{Ohio State University, Columbus, OH 43210}
\author{S. P. Rudin}
\affiliation{Theoretical Division, Los Alamos National Laboratory, Los Alamos, NM 87545}
\author{R. C. Albers}
\affiliation{Theoretical Division, Los Alamos National Laboratory, Los Alamos, NM 87545}
\author{J. W. Wilkins}
\affiliation{Ohio State University, Columbus, OH 43210}

\date{\today}

\begin{abstract}
For a previously published study of the titanium hcp ($\alpha$) to
omega ($\omega$) transformation, a tight-binding model was developed
for titanium that accurately reproduces the structural energies and
electron eigenvalues from all-electron density-functional
calculations.  We use a fitting method that matches the correctly
symmetrized wavefuctions of the tight-binding model to those of the
density-functional calculations at high symmetry points.  The
structural energies, elastic constants, phonon spectra, and
point-defect energies predicted by our tight-binding model agree with
density-functional calculations and experiment.  In addition, a
modification to the functional form is implemented to overcome the
``collapse problem'' of tight-binding, necessary for phase
transformation studies and molecular dynamics simulations.  The
accuracy, transferability and efficiency of the model makes it
particularly well suited to understanding structural transformations
in titanium.
\end{abstract}


\maketitle

\section{Introduction}

Titanium is a useful starting material for many structural
alloys;\cite{titanium03} however, the formation of the high-pressure
omega phase is known to lower toughness and ductility.\cite{Sikka82}
The atomistic mechanism of the transformation from the room
temperature $\alpha$ phase (hcp) to the high-pressure $\omega$ was
recently elucidated by Ref.~[\onlinecite{Trinkle03}].  The explication
of the \atow ~atomistic transformation relied on the comparison of
approximate energy barriers for nearly 1000 different 6- and 12-atom
pathways.  That study required the use of an accurate and efficient
interatomic potential model: in this case, a tight-binding model
reparameterized using all-electron density-functional calculations.

After reparameterizing, we modify the functional form of tight-binding
for small interatomic distances to overcome the collapse problem.
This ensures that the potential is suitable for phase transition
studies and molecular dynamics simulations.  The collapse problem for
tight-binding models is caused by unphysically large overlap at small
distances creating a low energy binding state; by modifying the
functional form using short-range splining, the collapse problem can
be avoided.  This paper provides the details of the model used in the
previous phase transformation study of Ref.~[\onlinecite{Trinkle03}] and
describes a general solution to the collapse problem.

Tight-binding is a parameterized electronic structure method for
calculation of total energies and atomic forces for arbitrary
structures.  It is an empirical model that can reproduce
density-functional results for a range of structures yet requires
orders of magnitude less computational effort.  The parameters of the
model are determined by fitting to a database and the range of
applicability is determined by comparison to structures not in the
database.  The end result is a model that balances three competing
properties---efficiency, accuracy, and transferability---which make it
applicable to a variety of important structures.

We fit our model to total energies and electron eigenvalues for
several crystal structures over a range of volumes to produce a
transferable model for the study of the \atow
~transformation.\cite{Trinkle03} Our fitting database is chosen to
sample a large portion of the available phase space of parameters
while constraining those parameters as much as possible.  The
resulting model accurately reproduces total energies, elastic
constants, phonons, and point defects; all of which are necessary for
transformation modeling.  In addition, the functional forms are
modified for small distances to overcome the unphysical collapse
problem; this is necessary for phase transitions and molecular
dynamics which sample small interatomic distances.

Section~\ref{sec:method} describes tight-binding as a parameterized
electronic structure method, the functional forms for titanium, the
modifications for short distances, our fitting database and our method
of optimization.  Section~\ref{sec:results} gives the optimized
parameters, and tests our model against total energies, elastic
constants, phonons, and point defect formation energies for $\alpha$,
$\omega$, and bcc Ti.  The point defect formation energies are used to
compare our parameters to those of Mehl and
Papaconstantopoulos\cite{NRL02} and Rudin~\et,\cite{Rudin04} and to
demonstrate the efficacy of our modification of the short-range
Hamiltonian and overlap functions.

%%%%%%%%%%%%%%%%%%%%%%%%%%%%%%%%%%%%%%%%%%%%%%%%%%%%%%%%%%%%%%%%%%%%%%%%
\section{Methodology}
\label{sec:method}

\subsection{Tight-binding formulation}
Electronic structure methods separate the total energy of a crystal
into an ionic contribution and an electronic contribution derived as
the solution to a Hamiltonian problem.  Treating electrons as
non-interacting fermionic quasiparticles permits an appropriate
one-particle solution.\cite{Koh65} To numerically solve the electronic
problem requires a set of basis functions $\phi_i$, in terms of which
the matrix $H_{ij}$ of the Hamiltonian operator and overlap matrix
$S_{ij}$ are
\beu
H_{ij} = \langle \phi_i | \hat H | \phi_j\rangle, 
S_{ij} = \langle \phi_i | \phi_j\rangle.
\eeu
These matrices give the eigenvalue equation,
\be
\label{eqn:eigenvalue}
H\psi_n = \epsilon_n S\psi_n,
\ee
where the electronic contribution to the total energy includes the
term 
\beu
2 \sum_{\epsilon_n<\EF} \epsilon_n,
\eeu
with Fermi energy $\EF$.  The Hamiltonian contains information about
the wavefunction solutions themselves (e.g., density-functional
theory).  Typically, the wavefunctions must be found
self-consistently, which increases the computational requirements.

In the tight-binding method, approximate Hamiltonian and overlap
matrices are constructed by assuming atom-centered orbitals in a
two-center approximation.  This technique is related to the linear
combination of atomic-like orbitals (LCAO) method, which uses a basis
$\phi_i$ of solutions to the isolated atomic Schr\"odinger equation up
to some energy and angular momentum quantum numbers $(nl)$:
$\phi_{nlm}(\vec r)=f_{nl}(|r|)Y_{lm}(\vec r/|r|)$.  Tight-binding
Hamiltonian and overlap functions are calculated independently of the
local environment which increases efficiency but at the expense of
transferability.

Empirical tight-binding eliminates explicit basis functions from the
problem and parameterizes the Hamiltonian and overlap matrices in
terms of simple two-center integrals.\cite{Sla54}  The basis is
chosen to be angular momentum solutions ${lm}$ up to some maximum $l$
value: For a maximum $l=1$ we use $s$, $p_x$, $p_y$ and $p_z$ as the
basis functions; for a maximum of $l=2$, we add in the five
$d$-orbitals $d_{xy}$, $d_{yz}$, $d_{zx}$, $d_{x^2-y^2}$ and
$d_{3z^2-r^2}$.  The Hamiltonian and overlap matrices are written as
sums of parameterized functions $\bar h_{lm,l'm'}(\vec r)$ and $\bar
s_{lm,l'm'}(\vec r)$ where $\vec r = \vec R_i - \vec R_{j}$ is the
separation between two atoms $i$ and $j$.  The two-center
approximation allows these functions to be simplified further
according to the angular momentum components of the basis.\cite{Sla54}
For example, $\bar h_{p_z,p_z}(\vec r)$ separates into two symmetrized
integrals
\beu
\bar h_{p_z,p_z}(\vec r) = h_{pp\sigma}(r)\cos^2\theta_z + 
h_{pp\pi}(r)\sin^2\theta_z,
\eeu
where $\theta_z$ is the angle between $\vec r$ and the $z$ axis.  The
higher rotational angular momentum integral $h_{pp\delta}(r)$ is zero
because a $p$ orbital has a maximal azimuthal quantum number of 1
along the $z$ axis.  The integrals $h_{pp\sigma}(r)$ and
$h_{pp\pi}(r)$ are functions of only the distance of separation
$r=|\vec R_j - \vec R_{j'}|$.  We write each Hamiltonian and overlap
integral in these symmetrized functions; for a model with an $spd$
basis, there are ten integrals (for $h$ and $s$) to be determined:
$(ss\sigma)$, $(sp\sigma)$, $(pp\sigma)$, $(pp\pi)$, $(sd\sigma)$,
$(pd\sigma)$, $(pd\pi)$, $(dd\sigma)$, $(dd\pi)$, and $(dd\delta)$.
The Hamiltonian and overlap matrices are then computed for an
arbitrary atomic arrangement.  The total energy of the system is
given by the eigenvalues of the Eq.~(\ref{eqn:eigenvalue})
$\epsilon_n$ and the ionic contribution:
\beu
E_{\mathrm{total}} = 2 \sum_{\epsilon_n<\EF} \epsilon_n + V(R_{\mathrm{nuclei}}),
\eeu
where $V$ does not depend on the electronic states of the system.

We use functional forms developed at the U.S. Naval Research
Laboratory (NRL) that do not use an explicit external pair potential
but instead has environment dependent onsite energies.\cite{NRL96,
NRL94, NRL98} The onsite Hamiltonian elements $\epsilon_s$,
$\epsilon_p$, and $\epsilon_d$ are not constants, but rather, depend
on the distances of neighboring atoms to approximate three-body
terms.%
\footnote{The energy of an orbital $l$ at atom $i$ in the
two-body approximation is $\langle \phi_{l,i} | \hat H
|\phi_{l,i}\rangle$; the first three-body correction is to include
hopping from atom $i$ to $j$ and back to $i$.}
The onsite energies $\epsilon_{l,i}$ are functions of the ``local
density'' $\rho_i$ with four parameters
\be
\label{eqn:onsite}
\epsilon_{l,i} = a_l + b_l\rho_i^{2/3} + c_l\rho_i^{4/3} + d_l\rho_i^2,
\ee
where
\be
\label{eqn:rho}
\rho_i = \sum_{j\ne i} \exp (-\lambda^2 r_{ij}) f_c (r_{ij}).
\ee
The smooth cutoff function $f_c(r)$ is
\be
\label{eqn:cutoff}
f_c(r) = \left( 1 + \exp\left(\frac{r-R_0}{l_0}\right) \right)^{-1}.
\ee
The intersite functions $h_{ll'm}(r)$ and $s_{ll'm}(r)$ are given by
three parameters each
\be
\label{eqn:hopping}
\begin{split}
h_{ll'm}(r) &= (e_{ll'm} + f_{ll'm}r)\exp (-g^2_{ll'm} r) f_c(r),\\
s_{ll'm}(r) &= (\bar e_{ll'm} + \bar f_{ll'm}r)\exp (-\bar g^2_{ll'm}
r) f_c(r).
\end{split}
\ee
The squared parameters $g_{ll'm}$ and $\bar g_{ll'm}$ guarantee the
exponential terms to decay with increased distance.

The overlap and Hamiltonian functions have an unfortunate behavior for
small distances $r$ which can lead to catastrophic failure in the
Hamiltonian problem.  The functional form in Eq.~(\ref{eqn:hopping})
is exponentially damped as $r$ grows; in reverse, this means that our
intersite functions {\em grow} exponentially as $r$ becomes small.  As
$h$ or $s$ between two atoms grow in magnitude they increase the
bonding between the two respective atoms; as $|s|\to 1$ the energy of
the bond grows as $1/(1-|s|)$.  When the bond energy grows, the
bonding state is populated while the antibonding state is not; this
results in a net attractive force between the two atoms.  As the
interatomic distance shrinks, the entire overlap matrix $S$ ceases to
be positive-definite, and the Hamiltonian problem of
Eqn.~(\ref{eqn:eigenvalue}) is no longer solvable.  This causes the
``collapse problem'' in molecular dynamics: two atoms come close to
each other and see a large attractive force that pulls them towards
each other until $S$ is not positive-definite.  In actuality, the
Hamiltonian problem is not meaningful even {\em before} $S$ is not
positive-definite, because the model predicts a bond with an
unphysically low energy.  In a real material, the growth in bonding is
counteracted by Coulumb repulsion: a two-electron term that is
not included in the tight-binding formalism.

\begin{figure}
\begin{center}
\epsfig{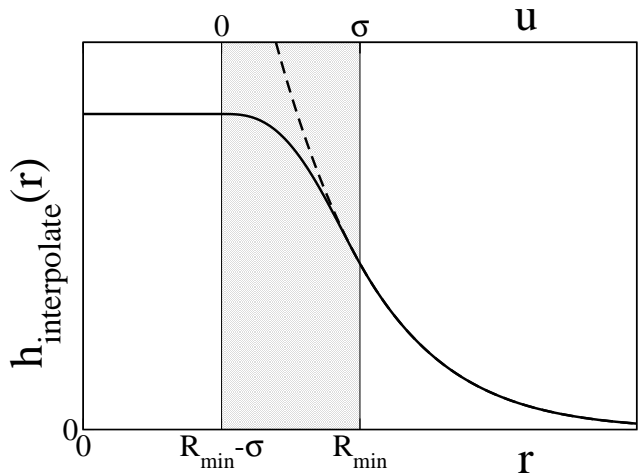}
\end{center}

\caption{Interpolated intersite function with short-range spline.  The
  parameterized function $h(r)$ grows exponentially as $r$ approaches
  zero, though the function is only sampled in the fitting database
  down to $\Rmin$.  At $r=\Rmin$, we replace the function with a
  quartic spline that matches the value, first and second derivatives
  at $\Rmin$; the dashed curve shows the growth of the original
  function.  The spline smoothly goes to a constant value in a width
  of $\sigma$.  Only one adjustable parameter, $\sigma$, is added to
  the entire fitting database, as $\sigma$ is the same for all
  functions.}
\label{fig:spline}
\end{figure}

{\em Short-range splining.}  To resolve this, we modify the intersite
functions to keep the overlap matrix $S$ positive definite.  Because
our fitting database includes only interatomic distances larger than
some minimum distance $\Rmin$, the functional form is guaranteed to be
correct only for $r>\Rmin$.  Below $\Rmin$, we smoothly interpolate
both $h_{ll'm}(r)$ and $s_{ll'm}(r)$ to a constant value.  The
interpolation is performed with a quartic spline, from $r=\Rmin$ down
to $r=\Rmin - \sigma$; below $\Rmin-\sigma$, the function takes on a
constant value.  We choose spline values to enforce continuity of
value and the first and second derivatives; the final functions for
both $h_{ll'm}(r)$ and $s_{ll'm}(r)$ are
\be
\label{eqn:spline}
h_\text{inter.}(r) = \begin{cases}
h(r)& : r>\Rmin,\\
h_\text{spline}(0)& : r<\Rmin-\sigma,\\
h_\text{spline}(r-\Rmin+\sigma)& : \mathrm{otherwise},\\
\end{cases}
\ee
where
\begin{multline}
h_\text{spline}(u) = h_0 -\frac{1}{2}\sigma h'_0 +
\frac{1}{12}\sigma^2 h''_0 + \left(\sigma h'_0 - \frac{1}{3}\sigma^2
h''_0\right)\frac{u^3}{\sigma^3} \\
+ \left(-\frac{1}{2}\sigma h'_0 
+ \frac{1}{4}\sigma^2 h''_0\right)\frac{u^4}{\sigma^4},
\end{multline}
for $u$ in $[0,\sigma]$, and $h_0$, $h'_0$, and $h''_0$ are the value,
first, and second derivative of $h(r)$ at $\Rmin$.
Figure~\ref{fig:spline} shows this interpolation schematically.  While
we smoothly interpolate $h_{ll'm}$ and $s_{ll'm}$, we \textit{retain}
the environment-dependent onsite terms; this has the effect of
reducing the strength of bonding while the onsite energy continues to
grow---effectively producing a pair repulsion between atoms at small
distances.

\begin{figure}
\begin{center}
\epsfig{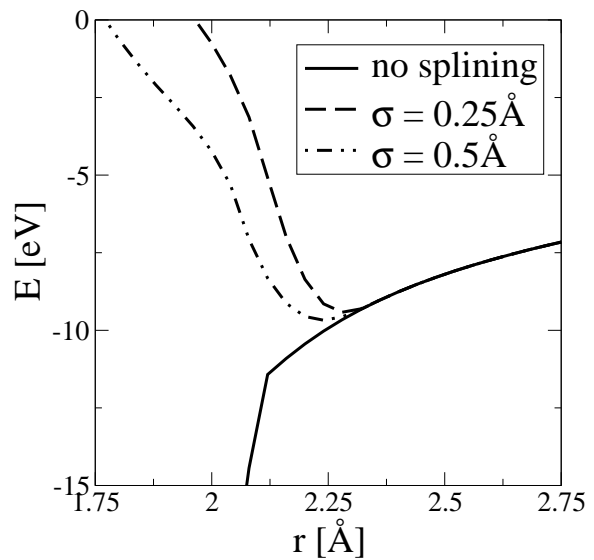}
\end{center}

\caption{Energy of Ti dimer calculated with tight-binding using
  short-range splining.  Without any short-range splining, the overlap
  matrix becomes artificially large, creating a bonding state with
  very low energy at small distances; at 1.92\AA, the tight-binding
  dimer Hamiltonian problem becomes unsolvable.  As described in the
  text, by short-range splining of the Hamiltonian and overlap
  functions, the model is stable and becomes repulsive at small
  distances.}
\label{fig:dimer}
\end{figure}

Figure~\ref{fig:dimer} illustrates the collapse problem for the Ti
dimer and how short-range splining stabilizes the model for small
distances.  As the distance between the two atoms decreases, a bonding
state with an artificially low energy decreases the dimer energy.  The
precipitous drop in the energy of this bonding state is due to an
increase in the overlap; at 1.92\AA, the overlap matrix becomes
non-positive definite, and the eigenproblem is no longer solvable.  A
$\sigma$ value of 0.5\AA\ or 0.25\AA\ makes the dimer stable; this is
necessary but not sufficient to solve the collapse problem for all
cases.

Our parameterization has 74 parameters to be optimized, plus 3 fixed
parameters.  The cutoff function $f_c(r)$ has two fixed parameters
$R_0$ and $l_0$, while the minimum distance $\Rmin$ is set by the
database.  There are 10 Hamiltonian and 10 overlap functions, each
with 3 parameters for a total of 60 parameters.  The 3 onsite energy
functions have 4 parameters each, and a single parameter $\lambda$ for
the density gives 13 parameters.  Finally, the short-range spline
range parameter $\sigma$ is determined using the dimer, and testing
with molecular dynamic calculations and defect relaxations.%
\footnote{Different values of $\sigma$ could be used for each
Hamiltonian and overlap element; we use a single parameter for
simplicity.}

\subsection{Fitting database}

We compile a database of electronic structure calculations of several
crystal structures using full-potential linearized augmented plane
wave (FLAPW) calculations\cite{Singh94} with the \wien\ program
suite.\cite{wien97} We use the generalized gradient approximation
(GGA) for the exchange-correlation energy.\cite{Perdew96} The sphere
radius is $\Rmt = 2.0\text{~bohr} = 1.06\text{~\AA}$; there is a
negligible charge leakage of $10^{-8}$ electrons.  The planewave
cutoff $\Kmax$ is given by $\Rmt\Kmax = 9$; this corresponds to an
energy cutoff of 275~eV.  The energy cutoff is not as large as
required in a typical pseudopotential calculation because the
planewaves are only used in the interstitial regions away from atom
centers.  The charge density is expanded in a Fourier series; the
largest magnitude vector in the expansion $G_\text{max}$ is
18~bohr$^{-1}$ (34~\AA$^{-1}$).  Local orbitals are used for the $s$,
$p$, and $d$ solutions inside the spheres.\cite{Singh94} Our core
configuration is Mg with semi-core $3p$ states represented by the
local $p$ orbitals; our $4s$, $3d$, and $4p$ states are the valence
orbitals.  A Fermi-Dirac smearing of 20~mRyd (272~meV) is used to
calculate the total energy.%
\footnote{The large smearing was used to reduce error introduced with
a small number of $k$-points; Ref.~[\onlinecite{Jones02}] used smaller
smearings without adverse effects.}

\begin{table}
\caption{Crystal structures used in tight-binding fitting database.
  Five different crystals structures are used, with five volumes for
  each of the cubic crystal structures.  The lattice constant a$_0$,
  volume per atom, and nearest neighbor distance and multiplicity for
  each structure is listed.  The equilibrium lattice constant for each
  structure is marked with an asterisk.  The same $k$-point mesh is used
  for all volumes of a given structure, and is constructed using the
  prescription of Ref.~\protect[\onlinecite{Chadi73,Monkhorst76}].
  The smallest distance to appear in this fitting database is
  $\Rmin=2.350\text{~\AA}$.}
\label{tab:fitdata}
  \begin{ruledtabular}
  \begin{tabular}{lcccl}
  \rentry{Structure} & 
  \rentry{a$_0$ (\AA)}&
  \rentry{V/atom (\AA$^3$)} &
  \rentry{nn (\AA)}&
  \rentry{$k$-point mesh}\\
bcc& 2.887 & 12.03 & 2.500 $\x 8$ & shifted $5\x 5\x 5$\\
& 3.060 & 14.32 & 2.650 $\x 8$ &  \quad (44 points)\\
& 3.281$^*$ & 17.66 & 2.841 $\x 8$ & \\
& 3.406 & 19.76 & 2.950 $\x 8$ & \\
& 3.579 & 22.93 & 3.100 $\x 8$ & \\
\hline
fcc& 3.747 & 13.16 & 2.650 $\x 12$ & unshifted $5\x 5\x 5$\\
 & 3.960 & 15.52 & 2.800 $\x 12$ & \quad (47 points) \\
 & 4.127$^*$ & 17.57 & 2.919 $\x 12$ \\
 & 4.384 & 21.06 & 3.100 $\x 12$ \\
 & 4.596 & 24.27 & 3.250 $\x 12$ \\
\hline
sc& 2.350 & 12.98 & 2.350 $\x 6$ & shifted $5\x 5\x 5$\\
 & 2.500 & 15.62 & 2.500 $\x 6$ & \quad (35 points)\\
 & 2.645$^*$ & 18.50 & 2.645 $\x 6$ \\
 & 2.800 & 21.95 & 2.800 $\x 6$ \\
 & 2.950 & 25.67 & 2.950 $\x 6$ \\
\hline
$\alpha$& 2.952$^*$ & 17.69 & 2.952 $\x 12$ & unshifted $5\x 5\x 2$\\
 & \multicolumn{3}{l}{($c/a=1.588$)} & \quad (42 points) \\
$\omega$& 4.600$^*$ & 17.23 & 2.656 $\x 3$ & unshifted $3\x 3\x 4$\\
&  \multicolumn{3}{l}{($c/a=0.613$)} & \quad (35 points)\\
& \multicolumn{4}{l}{$^*$ FLAPW equilibrium lattice constant}
  \end{tabular}
  \end{ruledtabular}
\end{table}

Table~\ref{tab:fitdata} shows a summary of the fitting database; it
consists of the total energies and eigenvalues on a $k$-point grid for
several crystal structures.  Five structures are used: simple cubic
(sc), body-centered cubic (bcc), face-centered cubic (fcc), hexagonal
closed-packed ($\alpha$), and omega ($\omega$).  The three cubic
structures are calculated over a range of volumes, while the hexagonal
structures are calculated only at FLAPW equilibrium volumes and $c/a$
ratios.  The eigenvalues in each structure are each shifted by a
constant amount so that the sum of the occupied bands (smeared by
272~meV) is the total energy.  We calculate the 9 lowest bands per
atom above the semicore $3p$ states; these represent the $4s$, $3d$
and $4p$ states both below and above the Fermi level.  We use the
lowest 6 bands at each $k$-point for fitting the cubic structures, 9
bands for $\alpha$, and 12 bands for $\omega$.

In addition to eigenvalues on a regular grid, we include eigenvalues
at high symmetry points and directions in the Brillouin zone to aid in
fitting.\cite{Papaconst86,Jones02} For the three cubic structures, we
calculate the eigenvalues at several high-symmetry points and
directions (10 for bcc and sc, and 12 for fcc) and then decompose the
electronic wavefunctions in terms of the symmetry character of the
eigenvalues.\cite{Cornwell69} Again, we use the lowest 6 states for
the high-symmetry points.  We are careful not to fit too many
eigenvalues at high-symmetry points, since the lowest 9 bands in the
GGA band structure may not correspond to those predicted in our $spd$
basis.%
\footnote{For example, in a bcc lattice the $\Gamma_{25}$ states are
  lower in energy than the $\Gamma_{15}$ states; however, the
  $\Gamma_{25}$ state corresponds to the $f$ orbitals $x(y^2-z^2)$
  etc., while the $\Gamma_{15}$ states correspond to $p$ orbitals $x$,
  $y$, $z$.  Hence, we only fit the lowest 6 states at $\Gamma$ to
  exclude the $\Gamma_{25}$ states from the fit.}

Because our fit includes the electron eigenvalues, we expect our model
to reproduce both total energies and energy derivatives.  Phonons and
elastic constants can be written in terms of the forces on atoms due
to small displacements; the Hellman-Feynman theorem relates the force
on an atom $R_i$ to the eigenvalues as
\beu
F_i = -2\sum_{\eps_n < \EF} \langle \psi_n| \frac{\partial \hat H}{\partial
R_i} |\psi_n\rangle.
\eeu
Thus, the electron eigenvalues of the bulk crystal contain
information about phonons and elastic constants.

\subsection{Optimization of parameters}

The parameters are optimized to minimize the mean squared error.  We
use the non-linear least-squares minimization method of
Levenberg-Marquardt with a numerical Jacobian.\cite{Marquardt63} We
weight each $k$-point by unity, and the resulting total energy by 200;
accordingly the total energies are weighted approximately the same as
the $k$-point data.  We initialize our parameters using the
Hamiltonian and overlap values for Ti from
Ref.~[\onlinecite{Papaconst86}] adapted to our functional form.  We then
fit only the environment-dependent onsite terms to the band-structure
of the cubic elements.  After an initial fit is found, we include the
hopping terms in the optimization.  We proceed using only the cubic
band structure, then the cubic band structure and total energies, and
finally all structures and energies.  After a new minimum is found, we
check each function to see if the minimization has made the
exponential term $g_{(ll'm)}$ too large; this corresponds to making
the entire function approximately zero over the sampled range of $r$
values.  We remedy this by resetting the $e$, $f$, and $g$ parameters
to 0, 0, and 0.5.  Several fitting runs are performed until the entire
fit set is accurately reproduced.

%%%%%%%%%%%%%%%%%%%%%%%%%%%%%%%%%%%%%%%%%%%%%%%%%%%%%%%%%%%%%%%%%%%%%%%%
\section{Results}
\label{sec:results}

\subsection{Parameters and fitting residuals}

\begin{figure}
\begin{center}
\epsfig{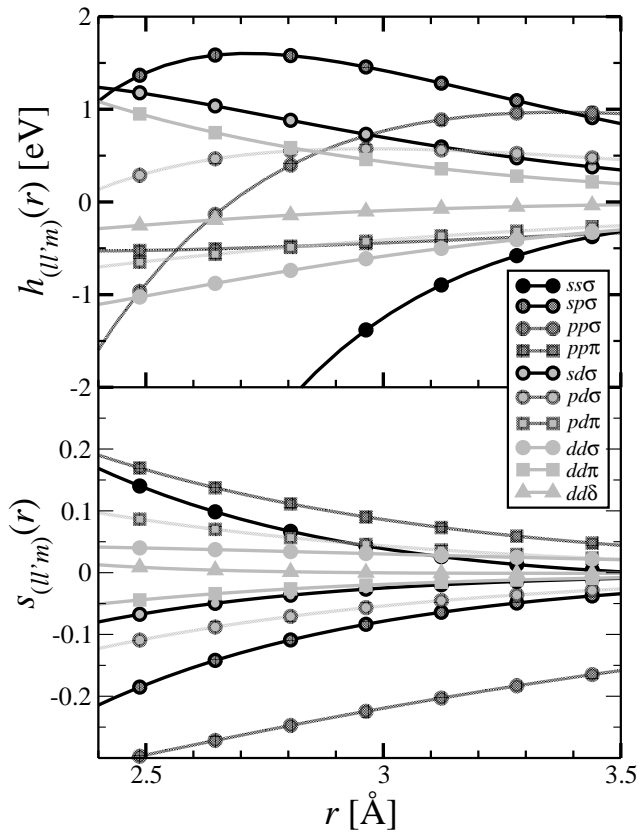}
\end{center}

\caption{Tight-binding intersite Hamiltonian and overlap
  functions.  The parameterized hopping integrals are shown for
  distances from 2.4~\AA\ to 3.5~\AA.  The $\Rmin$ in the fit is
  2.35~\AA; below this, these functions are smoothly interpolated to a
  constant value.  Circles represent $\sigma$ integrals, squares $\pi$
  integrals, and triangles $\delta$ integrals; black is for $s$, dark
  gray for $p$, and light gray for $d$.}
\label{fig:param}
\end{figure}

\begin{figure}
\begin{center}
\epsfig{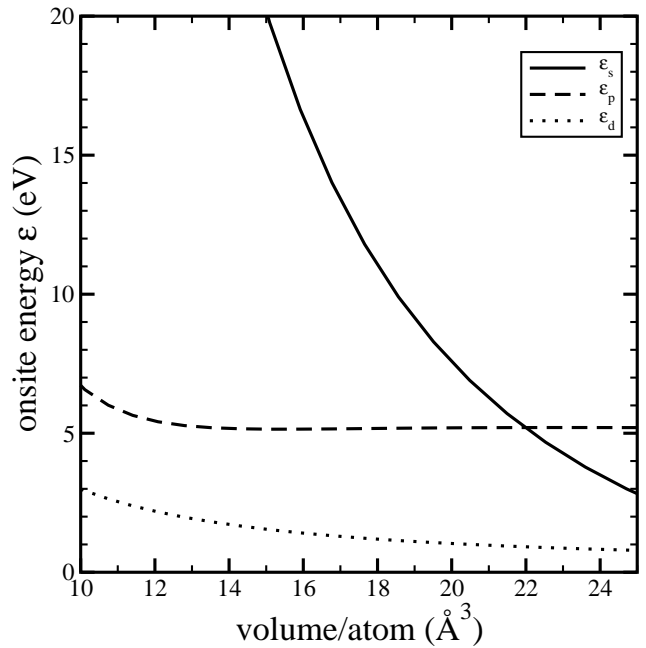}
\end{center}

\caption{Tight-binding onsite energy terms for hcp structure.  The
  onsite energies are environment dependent in our model; we show the
  variation with respect to the volume of an hcp crystal with
  $c/a=1.588$.  The low volume of 10~\AA$^3$ has a lattice constant of
  2.44~\AA, and the high volume of 25~\AA$^3$ has a lattice constant
  of 3.31~\AA.  The equilibrium hcp volume is 17.56~\AA$^3$.}
\label{fig:onsite}
\end{figure}

Table~\ref{tab:param} lists the parameters of the optimized
tight-binding model.  Figure~\ref{fig:param} shows the hopping
integrals $h_{(ll'm)}(r)$ and $s_{(ll'm)}(r)$ for a range of volumes;
the $\Rmin$ in the database is 2.35~\AA, and we interpolate each
function to a constant value below $\Rmin$.  Finally,
Figure~\ref{fig:onsite} shows the environment-dependent onsite energies
as a function of volume for an hcp crystal with $c/a=1.588$.

\newcommand*{\EQNNUM}[1]{\multicolumn{1}{r}{(\protect\ref{#1})}}
\begingroup
\squeezetable
\begin{table}
\caption{Tight-binding parameterization for titanium.  The onsite
  parameters are given for the $s$, $p$, and $d$ orbitals.  Each term
  is density dependent; the parameter in the density dependence is
  $\lambda$.  The cutoff function has fixed parameters $R_0$ and
  $l_0$.  Next, the intersite Hamiltonian and overlap elements are
  given for each of the 10 symmetrized $(ll'm)$ combinations.  Below
  $\Rmin$, each intersite function is smoothly interpolated to a
  constant value over the range $\sigma$.}
\label{tab:param}

%  \begin{ruledtabular}
  \begin{tabular}{rcccc}
  \hline\hline
& \rentry{$a_l$ (eV)} & \rentry{$b_l$ (eV)}
& \rentry{$c_l$ (eV)} & \rentry{$d_l$ (eV)} \\
$s$: & $-3.272\EE{0}$ & $3.714\EE{2}$ & $8.029\EE{3}$ & $7.879\EE{4}$\\
$p$: & $4.974\EE{0}$ & $3.747\EE{1}$ & $-1.874\EE{3}$ & $2.721\EE{4}$\\
$d$: & $3.632\EE{-1}$ & $3.238\EE{1}$ & $8.877\EE{1}$ & $9.355\EE{2}$\\
\multicolumn{4}{c}{$\epsilon_{l,i} = a_l + b_l \rho_i^{2/3} + c_l
\rho_i^{4/3} + d_l \rho_i^2$}&\EQNNUM{eqn:onsite}
  \end{tabular}

  \begin{align*}
    \rho_i = \sum_{j\ne i} \exp(-\lambda^2 r_{ij}) f_c(r_{ij})
      \text{\,:\quad} 
    &\lambda^2 = \left(0.3620\text{~\AA}\right)^{-1}
    &\text{(\protect\ref{eqn:rho})} \\
    f_c(r) = \left( 1 + \exp\left(\frac{r-R_0}{l_0}\right)\right)^{-1}
      \text{\,:\quad}
    &\ba{c}R_0 = 6.615\text{~\AA}\\l_0 = 0.2646\text{~\AA}\ea
    &\text{(\protect\ref{eqn:cutoff})}
  \end{align*}

  \begin{tabular}{rcccc}
& & \rentry{$e_{(ll'm)}$} & \rentry{$f_{(ll'm)}$}
  & \entry{\raisebox{3pt}[\depth][\width]{1/$g_{(ll'm)}^{2}$}} \\
$ss\sigma$: & h= & $-1.086\EE{2}$ eV & $-3.900\EE{3}$ eV/\AA & $0.3277$ \AA\\
 & s= & $9.277$ & $-2.624$ \AA$^{-1}$ & $0.8357$ \AA\\
\hline
$sp\sigma$: & h= & $-1.793\EE{3}$ eV & $8.066\EE{2}$ eV/\AA & $0.4926$ \AA\\
 & s= & $-11.81$ & $0.02523$ \AA$^{-1}$ & $0.5993$ \AA\\
$pp\sigma$: & h= & $-4.865\EE{2}$ eV & $1.816\EE{2}$ eV/\AA & $0.6929$ \AA\\
 & s= & $0.08093$ & $-1.351$ \AA$^{-1}$ & $1.036$ \AA\\
$pp\pi$: & h= & $1.202\EE{1}$ eV & $-8.252\EE{0}$ eV/\AA & $0.8925$ \AA\\
 & s= & $4.478$ & $-0.2899$ \AA$^{-1}$ & $0.8026$ \AA\\
\hline
$sd\sigma$: & h= & $-5.537\EE{2}$ eV & $3.096\EE{2}$ eV/\AA & $0.4772$ \AA\\
 & s= & $-4.331$ & $-5.085$ \AA$^{-1}$ & $0.4498$ \AA\\
$pd\sigma$: & h= & $-2.338\EE{2}$ eV & $9.994\EE{1}$ eV/\AA & $0.6321$ \AA\\
 & s= & $0.02557$ & $-3.383$ \AA$^{-1}$ & $0.5728$ \AA\\
$pd\pi$: & h= & $-4.979\EE{0}$ eV & $7.855\EE{-1}$ eV/\AA & $1.617$ \AA\\
 & s= & $0.1943$ & $2.308$ \AA$^{-1}$ & $0.5882$ \AA\\
\hline
$dd\sigma$: & h= & $1.706\EE{2}$ eV & $-1.150\EE{2}$ eV/\AA & $0.5266$ \AA\\
 & s= & $-0.9905$ & $0.7605$ \AA$^{-1}$ & $0.7990$ \AA\\
$dd\pi$: & h= & $9.920\EE{0}$ eV & $3.538\EE{1}$ eV/\AA & $0.5366$ \AA\\
 & s= & $-1.490$ & $-1.498$ \AA$^{-1}$ & $0.5213$ \AA\\
$dd\delta$: & h= & $1.109\EE{3}$ eV & $-6.205\EE{2}$ eV/\AA & $0.3340$ \AA\\
 & s= & $15.58$ & $-5.276$ \AA$^{-1}$ & $0.4412$ \AA\\
\multicolumn{5}{c}{$\{h,s\}_{(ll'm)}(r) = \left(e_{(ll'm)} +
f_{(llm')}r\right)\exp(-g^2_{(ll'm)}r)f_c(r)$~\text{(\protect\ref{eqn:hopping})}}\\
%% \EQNNUM{eqn:hopping}
\multicolumn{4}{c}{$\Rmin = 2.350\text{~\AA},\quad \sigma =
  0.265\text{~\AA}$} &\EQNNUM{eqn:spline}\\
  \hline\hline
  \end{tabular}
%  \end{ruledtabular}
\end{table}
\endgroup

To use the potential for phase-transformation studies, $\sigma$ was
determined by testing the stability of (1) the dimer, (2) molecular
dynamics runs, and (3) defect relaxations.  While the lowest energy
pathways studied by Ref.~[\onlinecite{Trinkle03}] have distances of closest
approach of 2.6~\AA, there were possible pathways where atoms
approached within 2.3~\AA\ of each other.  Without short-range
splining, calculations of energies of structures with distances below
our $\Rmin$ value can become problematic.  Initially, a $\sigma$ value
of 0.529~\AA\ was chosen based on the dimer; however, defect
relaxation calculations showed that a value of 0.265~\AA\ was
necessary to ensure stability for some of the point defects.

\newlength{\tableskip}
\setlength{\tableskip}{0pt}

\begin{table}
\caption{Fitting errors in total energy and $k$-points for tight-binding
  model.  For each structure, we report the absolute error in the
  total energy (first line) and the RMS error in all $k$-point energies
  in the fit set (second line).  The total energy errors are on the
  order of 1~meV, while the RMS band-structure errors are on the order
  of 100~meV.}
\label{tab:error}

  \begin{ruledtabular}
  \begin{tabular}{lccc}
 & \rentry{low volume} & \rentry{equilibrium} & \rentry{high volume} \\
bcc & 1.64~meV & 0.957~meV & 4.31~meV \\
 & 200~meV & 104~meV & 110~meV \\[\tableskip]
fcc & --1.79~meV & 1.25~meV & --0.821~meV \\
 & 136~meV & 87.1~meV & 114~meV \\[\tableskip]
sc & --0.0190~meV & --0.115~meV & --1.60~meV \\
 & 435~meV & 195~meV & 140~meV \\[\tableskip]
$\alpha$ & & --1.66~meV \\
 & & 69.1~meV \\[\tableskip]
$\omega$ & & --0.00993~meV \\
 & & 67.9~meV
  \end{tabular}
  \end{ruledtabular}
\end{table}

Table~\ref{tab:error} lists the errors in our tight-binding model with
respect to the fitting database.  Our average total energy errors are
approximately 1~meV; root-mean square errors in the $k$-point energies
are approximately 100~meV.  The tight-binding parameterization
adequately reproduces the database energetics.  To test
transferability, we compare to properties outside of this database.

\subsection{Total energies}

\begin{figure}
\begin{center}
\epsfig{file=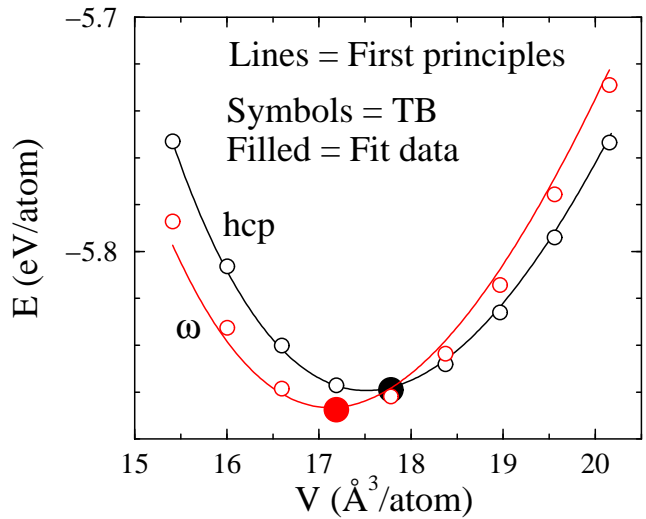, width=3.3in}
\end{center}

\caption{Comparison of tight-binding energy as a function of volume
  for $\alpha$ and $\omega$ with first principles data.  The two
  filled points were included in the fit; the lines are FLAPW total
  energies.  Our tight-binding model reproduces the fit
  data---slightly lower ground-state energy and equilibrium volume for
  $\omega$---and the equation of state of the full-potential
  calculations.}
\label{fig:bec}
\end{figure}

Figure~\ref{fig:bec} shows the tight-binding total energy as a
function of volume for $\alpha$ and $\omega$.  These curves were not
included in the fitting database; only the two points indicated.  We
reproduce both the slightly lower energy of $\omega$ over $\alpha$
predicted by pseudopotential methods\cite{Trinkle03} and FLAPW
calculations, as well as the slightly lower equilibrium volume of
$\omega$.  The three cubic structures were included in the fit and
have errors on the order of 3~meV/atom (c.f., Table~\ref{tab:error}).
This shows a wide range of applicability for our model under pressure.

\subsection{Elastic constants and phonons}
\label{app:phonons}

\begin{table}
\caption{Lattice parameters and elastic constants in units of GPa for
  $\alpha$, $\omega$, and bcc Ti from tight-binding, GGA, and
  experiment.  GGA corresponds to the elastic constants found using
  \vasp.\cite{Kresse93,Kresse96b} The experimental $\alpha$ elastic
  constants are measured at 4K,\cite{Simmons} and the bcc elastic
  constants at 1238K.\cite{Petry91} Our tight-binding model reproduces
  the GGA elastic constant combinations that preserve the symmetry of
  the structure (e.g., $C_{11}+C_{12}$), but has larger error with
  those that break it (e.g., $C_{44}$).  The deviation between the bcc
  experimental elastic constants and our calculations is due to the
  high temperature needed to stabilize the bcc structure in Ti.}
\label{tab:cij}

\begin{ruledtabular}
\begin{tabular}{lccccccc}
&\rentry{a (\AA)}&\rentry{c (\AA)}&\rentry{$C_{11}$}
&\rentry{$C_{12}$}&\rentry{$C_{13}$}&\rentry{$C_{33}$}&\rentry{$C_{44}$}\\
& \multicolumn{7}{c}{\hrulefill\ Tight-binding \hrulefill} \\
$\alpha$ & 2.94 & 4.71 &  155 & 91 & 79 & 173 & 65 \\
$\omega$ & 4.58 & 2.84 &  184 & 90 & 52 & 261 & 100 \\
bcc      & 3.27 & ---  &  87 & 112 & --- & ---  & 31 \\
& \multicolumn{7}{c}{\hrulefill\ GGA \hrulefill} \\
$\alpha$ & 2.95 & 4.68 &  172 & 82 & 75 & 190 & 45 \\
$\omega$ & 4.59 & 2.84 &  194 & 81 & 54 & 245 & 54 \\
bcc      & 3.26 & ---  &  95 & 110 & --- & --- & 42 \\
& \multicolumn{7}{c}{\hrulefill\ Experiment \hrulefill} \\
$\alpha$ & 2.95 & 4.68 &  176 & 87 & 68 & 191 & 51 \\
bcc      & 3.31 & ---  &  134 & 110 & --- & --- & 36
\end{tabular}
\end{ruledtabular}
\end{table}

Table~\ref{tab:cij} shows the equilibrium lattice constants and
elastic constants for $\alpha$, $\omega$, and bcc for our
tight-binding model.  The GGA numbers correspond to the elastic
constants found using \vasp.\cite{Kresse93,Kresse96b} Elastic constant
combinations which do not break symmetry such as $C_{11}+C_{12}$,
$C_{13}$, $C_{33}$ in the hexagonal crystals, and $C_{11}+2C_{12}$ in
bcc are reproduced within approximately 10\%.  However, the symmetry
breaking elastic constant combinations such as $C_{11}-C_{12}$ and
$C_{44}$ have larger errors.  It is worth noting that none of this
data, except for the bulk modulus of bcc, appears in any form in the
fitting database; the agreement is a consequence of reproducing the
electron eigenvalues.

We calculate phonons using the direct-force
method.\cite{Kunc82,Wei92,Frank95,Parlinski97} We calculate the forces
on all atoms in a supercell where one atom at the origin is displaced
by a small amount.  The numerical derivative of the forces with
respect to the displacement distance approximates the force constants
folded with the translational symmetry of the supercells.  The Fourier
transform of the force constants gives the dynamical matrix, and its
eigenvalues give the phonon frequencies.\cite{Ashcroft76} For $q$ vectors
commensurate with the supercell, the phonon frequencies are exact; for
incommensurate $q$ vectors, the calculated phonons are a Fourier
interpolation between exact values.  Our supercells are $4\x4\x3$ for
$\alpha$, $3\x3\x4$ for $\omega$, and $4\x4\x4$ simple cubic cell
for bcc; in all cases, a $2\x2\x2$ $k$-point mesh is used in the
supercell.

\newlength{\phononwidth}
\setlength{\phononwidth}{3.3in}

\begin{figure}
\begin{center}
\epsfig{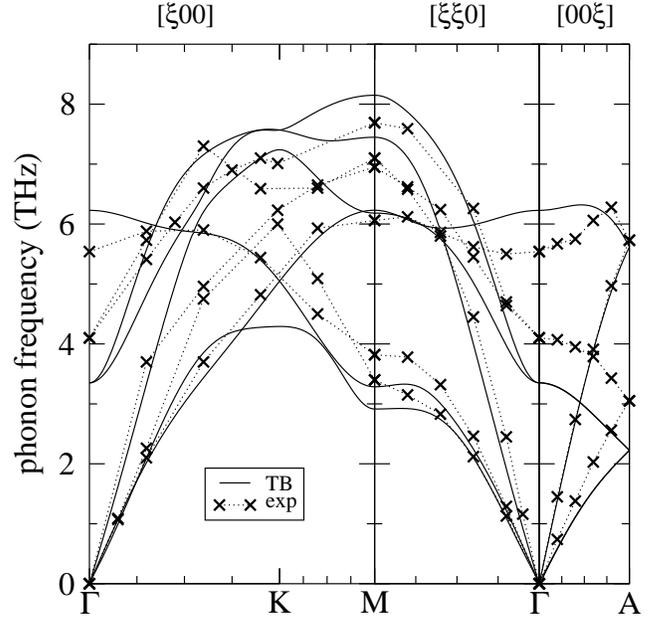}
\end{center}

\caption{Comparison of tight-binding phonons for the $\alpha$ phase
  with experimental phonon data.  The crosses are the experimental
  phonon frequencies at 295K.\protect\cite{Stassis79} The deviation
  from the experimental values at small $q$ corresponds to the
  mismatch in the $\alpha$ elastic constants.  Our tight-binding model
  does well for the high-energy optical and acoustic branches which
  are important for modeling the \atow ~transformation.}
\label{fig:hcp-phonons}
\end{figure}

\begin{figure}
\begin{center}
\epsfig{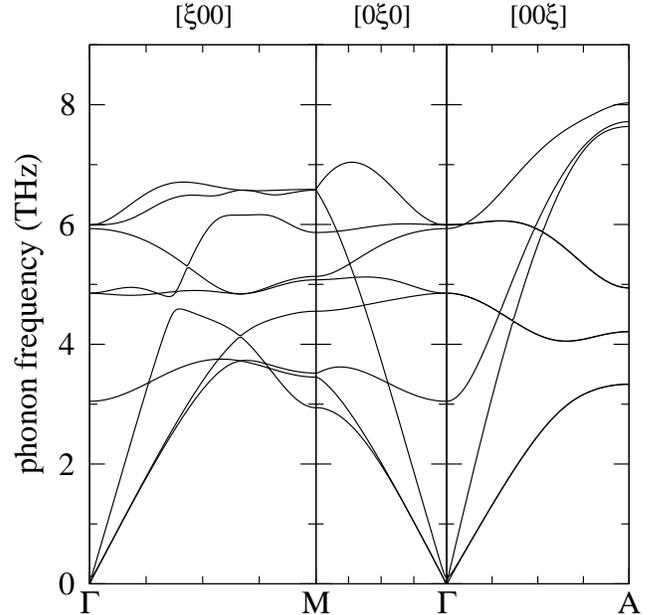}
\end{center}

\caption{Predicted $\omega$ phonons from tight-binding.  As expected
  from the $c/a$ ratio of 0.620, the phonon modes are stiffer along
  the $[00\xi]$ direction than the basal plane directions $[\xi00]$
  and $[0\xi0]$.}
\label{fig:omega-phonons}
\end{figure}

\begin{figure}
\begin{center}
\epsfig{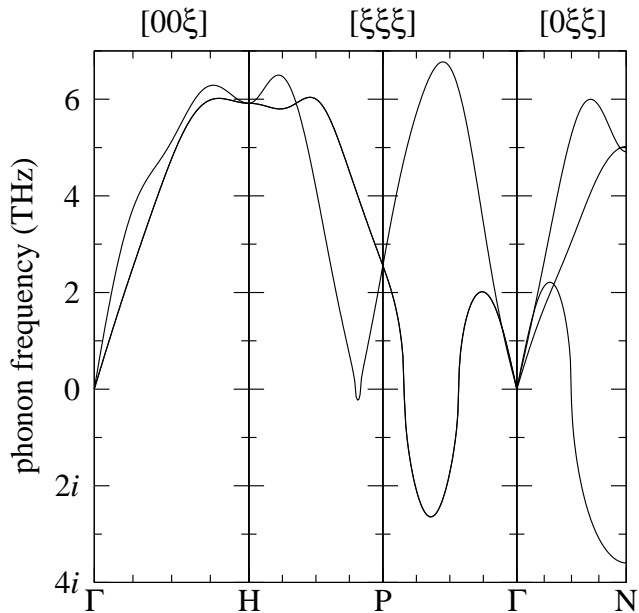}
\end{center}

\caption{Predicted bcc phonons from tight-binding.  At $T=0$, the bcc
  phase in Ti is unstable, as shown by the imaginary phonon
  frequencies.  The dip in the $[\xi\xi\xi]$ branch is near the
  L-$\frac{2}{3}[111]$ phonon, which corresponds to the
  bcc$\to$$\omega$ transformation pathway.  The imaginary phonon for
  T-$[011]$ corresponds to the bcc$\to$$\alpha$ transformation
  mechanism.\cite{Burgers34}}
\label{fig:bcc-phonons}
\end{figure}

Figures~\ref{fig:hcp-phonons}, \ref{fig:omega-phonons}, and
\ref{fig:bcc-phonons} are the predicted phonon dispersions for our
tight-binding model, calculated at the equilibrium volumes for each
structure.  The $\alpha$ phonons match the experimental values well
for the high energy phonons optical and acoustic branches; these are
important for modeling the shuffle during martensitic transformation.
The deviation from experiment for small $q$ corresponds to our
mismatch in elastic constants.  The $\omega$ phonons are expectedly
stiffer along the $c$ axis than in the basal plane due to the low
$c/a$ ratio.  The bcc phonons show phonon instabilities corresponding
to the bcc$\to$$\omega$ transformation (L-$\frac{2}{3}[111]$ phonon)
and the bcc$\to$$\alpha$ transformation (T-$[011]$
branch).\cite{Burgers34}

\subsection{Point defects}

Table~\ref{tab:defects} shows the formation energies of point defects
for $\alpha$ and $\omega$ at the equilibrium volumes for our
tight-binding model.  All $\alpha$ calculations are performed with a
$4\x4\x3$ (96 atom) supercell and all $\omega$ with a $3\x3\x4$ (108
atom) supercell.  No point defect information is included in the
initial fit; we reproduce the GGA formation energies for all of the
point defects considered.  This indicates that our tight-binding model
is applicable to the study of the \atow\ transformation path, where
atoms move out of their equilibrium configurations and often close to
one another.

\begin{table}
\caption{Point defect energies in eV for $\alpha$ and $\omega$ Ti from
  tight-binding for different parameterizations.  GGA refers to the
  defect formation energies calculated with \vasp; TB to the
  parameterization in this work; NRL to the parameterization by Mehl
  and Papaconstantopoulos;\protect\cite{NRL02} and LANL to the
  parameterization by Rudin~\et\protect\cite{Rudin04} NN refers to the
  distance of closest approach for two atoms in each defect in TB.
  The formation energies are calculated after relaxation.  The defects
  marked \textit{coll.} fell victim to the ``collapse problem'' during
  relaxation.  The $\alpha$-tetrahedral site is unstable, relaxing to
  form a dumbbell along the $[0001]$ direction in GGA.  The
  $\omega$-hexahedral site is very close to the $\omega$-tetrahedral
  site.\protect\cite{Hennig05} Many of the interstitial defects sample
  small distances, requiring the use of short-range splining to
  stabilize the defects.}
\label{tab:defects}

\begin{ruledtabular}
\begin{tabular}{lcccc|c}
\rentry{Defect}&\rentry{GGA}&\rentry{TB}&
  \rentry{NRL}&\rentry{LANL}&\rentry{NN [\AA]}\\
\multicolumn{6}{c}{\hrulefill\ $\alpha$ defects \hrulefill} \\
 Octahedral      & 2.58 & 2.89 & 1.31 & 2.55 & 2.50~\AA\\
 Tetrahedral     &\multicolumn{5}{c}{unstable} \\
 Dumbbell-$[0001]$ & 2.87 & 2.81 & 1.81 & \textit{coll.} & 2.18~\AA\\
 Vacancy         & 2.03 & 1.88 & 1.51 & 1.92 & 2.83~\AA\\
 Divacancy-AB    & 3.92 & 3.83 & 3.73 & 3.68 & 2.81~\AA\\
\multicolumn{6}{c}{\hrulefill\ $\omega$ defects \hrulefill} \\
 Octahedral      & 3.76 & 4.11 & 3.20 & 3.67 & 2.30~\AA\\
 Tetrahedral     & 3.50 & 3.58 & 2.86 & \textit{coll.} & 2.21~\AA\\
 Hexahedral      & 3.49 & 3.86 & 2.88 & 4.37 & 2.28~\AA\\
 Vacancy-A       & 2.92 & 2.85 & 2.99 & 3.25 & 2.60~\AA\\
 Vacancy-B       & 1.57 & 1.34 & 1.01 & 1.90 & 2.62~\AA
\end{tabular}
\end{ruledtabular}
\end{table}

The formation energies of point defects shows some improvement of our
model over two existing models.\cite{NRL02,Rudin04} The potential by
Rudin~\et\ uses the same functional forms as our potential without
short-range splining for hopping and overlap functions; Mehl and
Papaconstantopoulos use the same onsite function form, but adds
additional quadratic parameters to the hopping and overlap functions
in Eqn.~(\ref{eqn:hopping}).  All three potentials use the same onsite
functional forms.  For all three potentials, the binding energies
versus volume, elastic constants, and phonons are similar, though
Rudin's more accurately captures the low frequency $\alpha$ phonons.
However, point defect formation energies are better predicted by our
tight-binding parameterization.

The short distances sampled by the point defects emphasize the need
for short-range splining of both the overlap and Hamiltonian
functions.  The collapse of two defects in Rudin~\et's model is due to
the growth of the overlap matrices; the lower energies predicted by
Mehl and Papaconstantopoulus could be due to overly large overlap
elements at short-distances as well.  Interstitial defects, like phase
transformation pathways, can sample interatomic distances smaller than
the smallest distance included in the fitting database; without
short-range splining, this can lead to artificially lower energies, or
even collapse.  Without short-range splining, all three tight-binding
parameterizations fail for the Ti dimer at small distances: 1.92~\AA
~for this work, 1.76\AA\ for Mehl and Papaconstantopoulos, and
1.28~\AA ~for Rudin~\et ~The use of short-range splines provides a
solution to the collapse problem for non-orthogonal tight-binding
models.

\section{Conclusion}
\label{sec:concl}

We present an accurate and transferable tight-binding model with
parameters determined by density-functional calculations.  It
reproduces structural energies with pressure, elastic constants,
phonons, and point defect energies.  By fixing the short-range
behavior of the potential, point defects can be accurately computed,
which allows the calculation of energy barriers for phase
transformation pathways.  The wide range of applicability makes it
particularly well suited to the study of martensitic phase
transformations, such as \atow;\cite{Trinkle03} and short-range
splines represent a solution to the potential collapse problem of
non-orthogonal tight-binding models.

\begin{acknowledgments}
DRT thanks Los Alamos National Laboratory for its hospitality and was
supported by a Fowler Fellowship at Ohio State University. This
research is supported by DOE grants DE-FG02-99ER45795 (OSU) and
W-7405-ENG-36 (LANL). Computational resources were provided by the
Ohio Supercomputing Center and NERSC.
\end{acknowledgments}

\end{document}